\documentclass[preprint,showpacs,preprintnumbers,eqsecnum,epsbox]{revtex4}
\usepackage{graphicx}
\usepackage{dcolumn}
\usepackage{bm}

\def\lesssim{\ \raise.3ex\hbox{$<$}\kern-0.8em\lower.7ex\hbox{$\sim$}\ }
\def\gesim{\ \raise.3ex\hbox{$>$}\kern-0.8em\lower.7ex\hbox{$\sim$}\ }
\font\scripti=cmmi7
\font\scriptscripti=cmmi5
\def\sib#1{\setbox0 = \hbox{\scripti #1}
  \kern-.02em\copy0\kern-\wd0
  \kern.04em\box0} 
\def\ssib#1{\setbox0 = \hbox{\scriptscripti #1}
  \kern-.02em\copy0\kern-\wd0
  \kern.04em\box0} 
\font\tenib=cmmib10 
\skewchar\tenib='177 \skewchar\tenib='177 \skewchar\tenib='177
\def\pbold#1{\setbox0 = \hbox{$ #1 $}
  \kern-.022em\copy0\kern-\wd0
  \kern.011em\copy0\kern-\wd0
  \kern.011em\copy0\kern-\wd0
  \kern.011em\copy0\kern-\wd0
  \kern.011em\box0} 
\begin{document}
\title{On the Fulde-Ferrell State in Spatially Isotropic Superconductors}
\author{Y. Ohashi}
\affiliation{Institute of Physics, University of Tsukuba, Tsukuba,
  Ibaraki 305, Japan}
\date{\today}
\begin{abstract}
Effects of superconducting fluctuations on the Fulde-Ferrell (FF) state are discussed in a spatially isotropic three-dimensional superconductor under a magnetic field. For this system, Shimahara recently showed that within the phenomenological Ginzburg-Landau theory, the long-range order of the FF state is suppressed by the phase fluctuation of the superconducting order parameter. [H. Shimahara: J. Phys. Soc. Jpn. {\bf 67} (1998) 1872, Physica B {\bf 259-261} (1999) 492] In this letter, we investigate this instability of the FF state against superconducting fluctuations from the microscopic viewpoint, employing the theory developed by Nozi\'eres and Schmitt-Rink in the BCS-BEC crossover field. Besides the absence of the second-order phase transition associated with the FF state, we show that even if the pairing interaction is weak, the shift of the chemical potential from the Fermi energy due to the fluctuations is crucial near the critical magnetic field of the FF state obtained within the mean-field theory.
\end{abstract}
\pacs{03.75.Ss, 03.75.-b, 03.70.+k}
\maketitle
%
{\it Introduction--}
Under a strong magnetic field, the Fulde-Ferrell (FF) state is known to be more stable than the ordinary BCS state in singlet superconductors.\cite{Fulde,Larkin,Takada,Shimahara,Samokhin} The FF state has a spatially oscillating order parameter of $\Delta({\bf r})\sim {\rm e}^{{\rm i}{\bf q}_{\rm FF}\cdot{\bf r}}$ and can exist above the Clogston limit where the ordinary BCS state no longer exists.\cite{Clogston} Since this state originates from the inequivalence of the Fermi surfaces of electrons with up- and down-spins under magnetic field, the FF state can also be induced by a molecular field in ferromagnetic superconductors. Indeed, the FF state in the magnetic superconductor RuSr$_2$RECu$_2$O$_8$ (RE=Y, Gd) was recently discussed.\cite{Pickett,Shimahara2,Kanegae}
\par
In this letter, we discuss the effects of superconducting fluctuations on
the FF state. In ordinary BCS superconductivity, the long-range order
is possible in three-dimensional systems, while in one- and
two-dimensional cases, it is absent at finite temperatures, as proved
by Hohenberg.\cite{Hohenberg,note1} Although Hohenberg's theorem
does not apply to the FF state, the FF state is also usually
believed to be stable at least in three-dimensional systems. However,
Shimahara recently pointed out that within the phenomenological
Ginzburg-Landau (GL) theory, even in three-dimensional systems,
the long-range order of the FF state is absent due to the effect of the 
phase fluctuation of the order parameter when the system is spatially isotropic.\cite{Shimahara3,Shimahara3b} In this letter, we present a microscopic
approach to the stability of the FF state, focusing on the phase
transition temperature $T_{\rm c}$. We show the absence of the second-order FF phase transition, which is consistent with the result obtained by Shimahara.\cite{Shimahara3,Shimahara3b} In addition, we point out the importance of the contribution of fluctuation to the chemical potential in the absence of the second-order phase transition associated with the FF state. 
\par
The theory describing the effect of superconducting fluctuations on
$T_{\rm c}$ within the Gaussian approximation was developed by
Nozi\'eres and Schmitt-Rink (NSR theory)\cite{Nozieres} to investigate
the BCS-BEC crossover problem in strong-coupling
superconductors. In this theory, $T_{\rm c}$ and the chemical
potential $\mu$ are determined self-consistently. The latter quantity
is determined from the equation of the number of electrons including
superconducting fluctuations. When the pairing interaction is strong,
$\mu$ deviates from the Fermi energy due to the fluctuations. On the
other hand, this effect is weak in weak-coupling BCS superconductivity. Thus in the latter regime, we can actually regard $\mu$ as the Fermi energy. 
\par
Based on the NSR theory, we can expect that if the FF state is unstable against superconducting fluctuations, the chemical potential plays an important role near the critical magnetic field of the FF state obtained within the mean-field theory ($h_{\rm FF}^{\rm MF}$) even in the weak-coupling regime. Motivated by this expectation, we apply the NSR theory to the FF state. Based on this framework, we show that near $h_{\rm FF}^{\rm MF}$, the shift of the chemical potential induced by the superconducting fluctuations is crucial in clarifying the absence of the second-order FF phase transition.
\par
\vskip2mm
{\it Application of the NSR theory to the FF state--} The NSR theory from the viewpoint of the Gaussian fluctuations was discussed by Engelbrecht et al. using the functional integral method.\cite{Randeria}
 In this letter, we extend their theory to a three-dimensional weak-coupling $s$-wave superconductor under a magnetic field. Although this extension is straightforward, the resulting fluctuation effect on the FF state is found to be quite different from the ordinary BCS case. 
\par
The partition function in the path integral representation is given by
$
Z=\int{\cal D}\Psi_\sigma^*{\cal D}\Psi_\sigma{\rm e}^{-S}$,
where $S$ is the action having the form 
\begin{equation}
S=\int_0^\beta{\rm d}\tau\int{\rm d}{\bf r}
\bigl[
\sum_\sigma
\Psi^*_\sigma
({\partial \over \partial\tau}+{{\hat {\bf p}}^2 \over 2m}-\mu-h\sigma)\Psi_\sigma
-U\Psi^*_\uparrow\Psi^*_\downarrow\Psi_\downarrow\Psi_\uparrow
\bigr].
\label{eq.2}
\end{equation}
Here, $\Psi({\bf r},\tau)$ is a Grassmann variable describing electrons. Electrons feel the external magnetic field (or molecular field) $h$ and pairing interaction $U$. In eq. (\ref{eq.2}), we have neglected the orbital effect associated with the magnetic field. Equation (\ref{eq.2}) also assumes a spatially isotropic system with no lattice effect.  
\par
We introduce the superconducting order parameter $\Delta$ using the Stratonovich-Hubbard transformation as
$
Z=
\int{\cal D}\Psi_\sigma^*{\cal D}\Psi_\sigma
{\cal D}\Delta^*{\cal D}\Delta
{\rm e}^{-{\tilde S}}\equiv
\int{\cal D}\Delta^*{\cal D}\Delta
{\rm e}^{-S_{\Delta}}
$
,
where
$
{\tilde S}=\int_0^\beta{\rm d}\tau\int{\rm d}{\bf r}
\bigl[
{|\Delta|^2 \over U}-{\hat \Psi}^\dagger{\hat G}^{-1}{\hat \Psi}
\bigr]
$.
In ${\tilde S}$, ${\hat \Psi}^\dagger\equiv(\Psi^*_\uparrow({\bf r},\tau),\Psi_\downarrow({\bf r},\tau))$, while ${\hat G}$ is given by
\begin{eqnarray}
{\hat G}^{-1}
&=&-({\partial \over \partial\tau}+h)
-({{\hat {\bf p}}^2 \over 2m}-\mu)\sigma_3
+
\left(
\begin{array}{cc}
0&\Delta({\bf r},\tau) \\
\Delta^*({\bf r},\tau) & 0
\end{array}
\right)
\nonumber
\\
&\equiv&
{\hat G}_0^{-1}+{\hat \Delta}
.
\label{eq.5}
\end{eqnarray} 
Here, $\sigma_i$ ($i=1,2,3$) are the Pauli matrices. 
\par
The mean field solution for the order parameter $\Delta_{\rm MF}$ is obtained from $\delta S_{\Delta}/\delta\Delta^*=0$, which gives the ordinary gap equation as $\Delta_{\rm MF}({\bf r})=U\langle\Psi_\downarrow({\bf r})\Psi_\uparrow({\bf r})\rangle$. When we consider the FF state described by $\Delta_{\rm MF}({\bf r})=\Delta_{\rm MF} {\rm e}^{{\rm i}{\bf q}_{\rm FF}\cdot{\bf r}}$, this equation can be written as
$
\Delta_{\rm MF}={U \over \beta}\sum_{{\bf p},\omega_m}
{\hat G}^{\rm MF}_{12}({\bf p},i\omega_m),
$ where ${\hat G}^{\rm MF}_{12}({\bf p},{\rm i}\omega_m)$ is the (12)-component of the one-particle matrix thermal Green function including the mean field $\Delta_{\rm MF}$:\cite{Takada}
\begin{eqnarray}
{\hat G}^{\rm MF}({\bf p},{\rm i}\omega_m)
=
\left(
\begin{array}{cc}
{\rm i}\omega_m-h-\varepsilon_{{\bf p}+{\bf q}_{\rm FF}/2}+\mu
&
\Delta_{\rm MF}
\\
\Delta_{\rm MF}^*
&
{\rm i}\omega_m-h+\varepsilon_{-{\bf p}+{\bf q}_{\rm FF}/2}-\mu
\end{array}
\right)^{-1}.
\label{eq,7}
\end{eqnarray}
Here, $\varepsilon_{\pm{\bf p}+{\bf q}_{\rm FF}/2}\equiv({\bf
  p}\pm{\bf q}/2)^2/2m$ is the kinetic energy while $\omega_m$ is the
fermion Matsubara frequency. At $T_{\rm c}$ with $h=h_{\rm FF}^{\rm
  MF}$, the gap equation is reduced to\cite{Fulde,Larkin,Takada,Shimahara}
\begin{equation}
1=U\sum_{\bf p}
{
1
-f(\varepsilon_{{\bf p}+{\bf q}_{\rm FF}/2}-\mu+h_{\rm FF}^{\rm MF})
-f(\varepsilon_{-{\bf p}+{\bf q}_{\rm FF}/2}-\mu-h_{\rm FF}^{\rm MF})
\over
\varepsilon_{{\bf p}+{\bf q}_{\rm FF}/2}+
\varepsilon_{-{\bf p}+{\bf q}_{\rm FF}/2}
-2\mu}
,
\label{eq.8}
\end{equation}
where $f(\varepsilon)$ is the Fermi distribution function. Within the mean field approximation, eq. (\ref{eq.8}) gives the well-known $h-T$ phase diagram shown in the inset in Fig. 1.\cite{Fulde,Takada}
\par
Now we consider the fluctuation effect around the mean field solution
in the normal state ($\Delta_{\rm MF}=0$). Executing the functional
integral in terms of $\Psi_\sigma$ and $\Psi_\sigma^*$ in the
partition function $Z$, we have
\begin{eqnarray}
S_\Delta
&=&
\int_0^\beta{\rm d}\tau\int{\rm d}{\bf r}
{|\Delta|^2 \over U}-
{\rm Tr}\log[-{\hat G}^{-1}]
\nonumber
\\
&=&
\int_0^\beta{\rm d}\tau\int{\rm d}{\bf r}
{|\Delta|^2 \over U}-
{\rm Tr}\log[-{\hat G}_0^{-1}]
-\sum_{n=1}^\infty{1 \over n}{\rm Tr}[-{\hat G}_0{\hat \Delta}]^n.
\label{eq.9}
\end{eqnarray}
In the following, we retain the summation in the last term up to $n=2$ (Gaussian approximation). The first-order term ($n=1$) vanishes because $\Delta_{\rm MF}=0$ above $T_{\rm c}$. Now, we write ${\hat \Delta}({\bf r},\tau)=\Delta^{(1)}\sigma_1-\Delta^{(2)}\sigma_2$ in eq. (\ref{eq.9}), where $\Delta^{(1)}$ and $\Delta^{(2)}$ are, respectively, the real and imaginary parts of $\Delta({\bf r},\tau)$. When we introduce the $({\bf p},\omega_m)$-representation instead of $({\bf r},\tau)$-representation, eq. (\ref{eq.9}) under the Gaussian approximation is reduced to ($q\equiv({\bf q},{\rm i}\nu_m)$, where $\nu_m$ is the boson Matsubara frequency)
\begin{eqnarray}
S_\Delta
&=&
-{\rm Tr}\log[-{\hat G}_0^{-1}]
+{1 \over \beta U}
\sum_q
(\Delta^{(1)}_{-q},\Delta^{(2)}_{-q})
\left(
\begin{array}{cc}
1+U\Pi_{11}(q)& -U\Pi_{12}(q)\\
-U\Pi_{21}(q)& 1+U\Pi_{22}(q)
\end{array}
\right)
\left(
\begin{array}{c}
\Delta^{(1)}_{q}\\
\Delta^{(2)}_{q}
\end{array}
\right).
\label{eq.10}
\end{eqnarray}
Here, $\Delta^{(j)}_q$ is the Fourier-transformed $\Delta^{(j)}({\bf r},\tau)$ and $\Pi_{ij}$ is the generalized polarization function defined by\cite{Ohashi}
$
\Pi_{ij}(q)={1 \over 2\beta}\sum_p
{\rm Tr}
\bigl[
G_0(p)\sigma_iG_0(p+q)\sigma_j
\bigr]
$,
where $G_0(p)^{-1}={\rm i}\omega_m-h-(\varepsilon_{\bf
  p}-\mu)\sigma_3$. In the superconducting state, when we take the
order parameter proportional to the $\sigma_1$-component, $\Pi_{11}$ and
$\Pi_{22}$ represent, respectively, the amplitude and phase
fluctuations of the order parameter, while $\Pi_{12}$ and $\Pi_{21}$
describe their coupling. In the normal state, they satisfy
$\Pi_{11}=\Pi_{22}$, $\Pi_{12}=-\Pi_{21}$, $\Pi_{jj}(-q)=\Pi_{jj}(q)$
and $\Pi_{12}(-q)=-\Pi_{12}(q)$. After the functional integration in
terms of $\Delta$ and $\Delta^*$ in the partition function $Z$, we obtain the
thermodynamic potential 
$
\Omega=-T\log Z
=\Omega_0+T\sum_q\log(1-U\Pi(q))
$
,
where $\Omega_0$ is the thermodynamic potential for free electrons; $\Pi(q)\equiv-(\Pi_{11}+{\rm i}\Pi_{22})$ describes superconducting fluctuations in the normal state:\cite{Nozieres}
\begin{eqnarray}
\Pi(q)
=
\sum_{\bf p}
{ 
1-
f(\varepsilon_{{\bf p}+{\bf q}/2}-\mu+h)-
f(\varepsilon_{{\bf p}+{\bf q}/2}-\mu-h)
\over 
\varepsilon_{{\bf p}+{\bf q}/2}+\varepsilon_{-{\bf p}+{\bf q}/2}-2\mu-{\rm i}\nu_m}
.
\label{eq.13}
\end{eqnarray}
When we take $h=0$, the thermodynamic potential $\Omega$ is reduced to that obtained by Nozi\'eres and Schmitt-Rink.\cite{Nozieres}
\par
The chemical potential is determined by the equation of the number of electrons, which is obtained from the identity $N=-\partial\Omega/\partial\mu$:
\begin{eqnarray}
N&=&
N_{\rm F}^0
+
{1 \over \beta}\sum_q 
{\rm e}^{{\rm i}\delta\nu_m}
{U \over 1-U\Pi(q)}
{\partial \Pi(q) \over \partial\mu}
\nonumber
\\
&=&
N_{\rm F}^0
+{1 \over \pi}
\sum_{\bf q}
\int_{-\infty}^\infty dz N_{\rm B}(z){\rm Im}
\bigl[
{U \over 1-U\Pi({\bf q},i\nu_m\to z+{\rm i}\delta)}
{\partial \Pi({\bf q},{\rm i}\nu_m\to z+{\rm i}\delta) 
\over 
\partial\mu}
\bigr]
\nonumber
\\
&\equiv&
N_{\rm F}^0+N_{\rm FL}.
\label{eq.14}
\end{eqnarray}
Here, $N_{\rm F}^0\equiv\sum_{{\bf p},\sigma}f(\varepsilon_{\bf p}-\sigma h)$ is the number of free electrons and $N_{\rm B}(z)$ is the Bose distribution function. In eq. (\ref{eq.14}), $N_{\rm FL}$ is the fluctuation contribution to $N$.
\par
For later convenience, we rewrite eq. (\ref{eq.8}) as
\begin{equation}
1=U\Pi({\bf q}_{\rm FF},{\rm i}\nu_m=0).
\label{eq.15}
\end{equation}
The transition temperature $T_{\rm c}$ and $\mu$ are determined self-consistently from eqs. (\ref{eq.14}) and (\ref{eq.15}). 
\par
\vskip2mm
{\it Stability of the FF state--}
In weak-coupling three-dimensional BCS superconductors, the fluctuation effect described by $N_{\rm FL}$ is weak; thus we can neglect $N_{\rm FL}$. Then, since eq. (\ref{eq.14}) is reduced to the equation of the number of free electrons, we can safely take $\mu=\varepsilon_{\rm F}$, where $\varepsilon_{\rm F}$ is the Fermi energy of free electrons at $h=0$, as long as the temperature and magnetic field dependence of $\mu$ can be neglected. 
\par
However, the situation is quite different in the FF case. At $h_{\rm FF}^{\rm MF}$, we find from eq. (\ref{eq.15}) that $\Pi({\bf q},0)$ must be maximum and be equal to $1/U$ at ${\bf q}_{\rm FF}$. As an example, we show in Fig. 1 the momentum dependence of $\Pi({\bf q},0)$ at $h_{\rm FF}^{\rm MF}(T/T^0_{\rm c}=0.1)$, where $T_{\rm c}^0$ is the phase transition temperature at $h=0$. We find that $\varepsilon_{\rm F}\Pi({\bf q},0)=2.5$, i.e., $U\Pi({\bf q},0)=1$, at $|{\bf q}|=0.024k_{\rm F}$, where $k_{\rm F}$ is the Fermi momentum. (For comparison, the case of $h_{\rm FF}^{\rm MF}(T/T^0_{\rm c}=0.7)$, at which the BCS state appears, is also shown. In this case, $\Pi({\bf q},0)$ is maximum at ${\bf q}=0$.) When we expand $\Pi({\bf q},{\rm i}\nu_m\to z+{\rm i}\delta)$ around ${\bf q}_{\rm FF}$ and $z=0$ at $h_{\rm FF}^{\rm MF}$, we obtain (note that $1-U\Pi({\bf q}_{\rm FF},0)=0$) 
\begin{equation}
1-U\Pi({\bf q},z)\simeq C({\bf q}_{\rm FF})(|{\bf q}|-|{\bf q}_{\rm FF}|)^2-{\rm i}\alpha z.
\label{eq.16}
\end{equation}
Here, $C({\bf q}_{\rm FF})=-U/2\cdot\partial^2\Pi({\bf q}_{\rm FF},0)/\partial q^2>0$ and $\alpha=(\pi U/8T)N(\mu){\rm sech}^2h/2T$, where $N(\mu)$ is the density of states at $\mu$.\cite{note3} We note that since the system is isotropic, the expansion in eq. (\ref{eq.16}) includes $(|{\bf q}|-|{\bf q}_{\rm FF}|)^2$, but not $({\bf q}-{\bf q}_{\rm FF})^2$. Substituting eq. (\ref{eq.16}) into eq. (\ref{eq.14}), we obtain 
\begin{eqnarray}
N&=&
N_{\rm F}^0
+{1 \over \pi}
\sum_{\bf q}
\int_{-\infty}^\infty dz N_{\rm B}(z)
{\rm Im}
[{U \over C({\bf q}_{\rm FF})
(|{\bf q}|-|{\bf q}_{\rm FF}|)^2-{\rm i}
\alpha z}
{\partial \Pi({\bf q},z) \over \partial\mu}]
\nonumber
\\
&\simeq&
N_{\rm F}^0
+{\alpha TU \over \pi}
{\partial \Pi({\bf q}_{\rm FF},0) \over \partial\mu}]
\sum_{\bf q}
\int_{-\infty}^\infty dz 
{1 \over C({\bf q}_{\rm FF})^2(|{\bf q}|-|{\bf q}_{\rm FF}|)^4+\alpha^2 z^2}
\nonumber
\\
&=&
N_{\rm F}^0
+{TU \over C({\bf q}_{\rm FF})}
{\partial \Pi({\bf q}_{\rm FF},0) \over \partial\mu}
\sum_{\bf q}
{1 \over (|{\bf q}|-|{\bf q}_{\rm FF}|)^2}.
\label{eq.17}
\end{eqnarray}
In the second line, since the region satisfying $z\sim 0$ and $|{\bf q}|\sim|{\bf q}_{\rm FF}|$ is important, we approximately set $N_{\rm B}(z)\simeq T/z$ and $\partial \Pi({\bf q},z)/\partial\mu\simeq\partial\Pi({\bf q}_{\rm FF},0)/\partial\mu$. The last expression in eq. (\ref{eq.17}) always diverges at finite temperatures as long as $|{\bf q}_{\rm FF}|\ne 0$, even in three-dimensional systems. Namely, no solution exists that satisfies both eqs. (\ref{eq.14}) and (\ref{eq.15}). Since the divergence comes from $N_{\rm FL}$, this result indicates that the superconducting fluctuations suppress the second-order phase transition associated with the FF state. 
\par
Besides the FF state, various nonuniform superconducting states which are described by $\Delta({\bf r})=\Delta\sum_j{\rm e}^{{\rm i}{\bf q}_j\cdot{\bf r}}$ ($|{\bf q}_j|={\rm const.}$) have been proposed under magnetic field (for example, $\Delta({\bf r})=\Delta\cos({\bf q}\cdot{\bf r})$).\cite{Larkin,Shimahara4} Although these states have different free energies below the critical magnetic field (within the mean-field theory), their second-order phase transitions are still determined by eqs. (\ref{eq.14}) and (\ref{eq.15}) in spatially isotropic systems. Thus, as in the case of the FF state, their second-order phase transitions are also absent due to the fluctuation effect.
\par
We also note that the factor $1/(|{\bf q}|-|{\bf q}_{\rm FF}|)^2$ in
eq. (\ref{eq.17}) leading to the divergence originates from the fact
that all the states satisfying $|{\bf q}|=|{\bf q}_{\rm FF}|$ are
degenerate in the present isotropic system. When we take into account
a crystal lattice, the band structure and/or the pairing
interaction may be anisotropic, which leads to the anisotropy of
$\Pi({\bf q},{\rm i}\nu_m)$ in momentum space. Then the direction of
the most stable FF state may become discrete, reflecting the crystal
symmetry. In this case, the factor $1/(|{\bf q}|-|{\bf q}_{\rm
  FF}|)^2$ in eq. (\ref{eq.17}) is replaced by $1/({\bf q}-{\bf
  q}_{\rm FF})^2$ around ${\bf q}_{\rm FF}$. Then, by the
transformation ${\bf q}-{\bf q}_{\rm FF}\to{\bar {\bf q}}$, the stability of the FF state becomes similar to that in the case of the ordinary BCS state; the divergence is absent
in three-dimensional systems. Namely, the FF state becomes stable in
spatially anisotropic three-dimensional systems where the direction of the most stable FF state is discrete.
\par
It is worth noting that the difference between the ordinary BCS state and the FF state against superconducting fluctuations originates from the spatially oscillating character of the latter's order parameter, $\Delta({\bf r})\sim \Delta {\rm e}^{{\rm i}{\bf q}_{\rm FF}\cdot{\bf r}}$. This wavy structure is similar to a "crystal of layers" stacking in the ${\bf q}_{\rm FF}$-direction, such as a smectic liquid crystal.\cite{note2} This type of crystal is known to be unstable against structural fluctuations in three-dimensional systems.\cite{Chaikin} Thus a similar situation is also expected in the FF state. Indeed, assuming ${\bf q}_{\rm FF}=(0,0,q_{\rm FF})$ and considering the fluctuations around ${\bf q}_{\rm FF}$, we find (${\bf q}_\perp\equiv(q_x,q_y)$, $\delta q_z\equiv q_z-q_{\rm FF}$)
\begin{eqnarray}
N_{\rm FL}
&\simeq&
{TU \over 2q_{\rm FF}|C({\bf q}_{\rm FF})|}
{\partial \Pi({\bf q}_{\rm FF},0) \over \partial\mu}
\sum_{\bf q}
{1 \over ({\bf q}^2-{\bf q}_{\rm FF}^2)^2}
\nonumber
\\
&\simeq&
\int {{\rm d}\delta q_z{\rm d}q_\perp \over 4\pi^2}
{q_\perp \over 4q_{\rm FF}^2(\delta q_z)^2+q_\perp^4},
\label{eq.18}
\end{eqnarray}
where we have dropped unimportant factors in the last expression. Equation (\ref{eq.18}) is identical to the expression for the effect of structural fluctuations of a layered crystal stacking in the $z$-direction in a three-dimensional system.\cite{Chaikin,Kleinert} This similarity clearly shows that "structural fluctuations" of the wavy structure of the FF order parameter are the key to the absence of the long-range order of this state. We also note that this kind of fluctuation effect also appears in a spiral magnetic state in magnetic superconductors.\cite{Kleinert}
\par
\vskip2mm
{\it Fluctuation contribution to $\mu$--}
Next we discuss what actually occurs near $h_{\rm FF}^{\rm MF}$, using eqs. (\ref{eq.14}) and (\ref{eq.15}). Near $h_{\rm FF}^{\rm MF}$, when $\Pi({\bf q},0)$ is maximum at ${\bf q}_{\rm max}$ ($\simeq{\bf q}_{\rm FF}$), eq. (\ref{eq.14}) can be evaluated as 
\begin{eqnarray}
N\simeq N_{\rm F}^0
+{TU}
{\partial \Pi({\bf q}_{\rm max},0) \over \partial\mu}
\sum_{\bf q}
{1 
\over 
(1-U\Pi({\bf q}_{\rm max},0))+
C({\bf q}_{\rm max})(|{\bf q}|-|{\bf q}_{\rm max}|)^2
}.
\label{eq.19}
\end{eqnarray}
As the magnetic field is lowered to approach $h_{\rm FF}^{\rm MF}$,
the factor $1-U\Pi({\bf q}_{\rm max},0)$ becomes small, so that the
fluctuation contribution to $N$ described by the second term in
eq. (\ref{eq.19}) ($N_{\rm FL}$) takes a large positive or negative value depending on the sign of $\partial \Pi({\bf q}_{\rm
  max},0)/\partial\mu$. When $\partial \Pi({\bf q}_{\rm
  max},0)/\partial\mu>0$, in order to satisfy eq. (\ref{eq.19}), $\mu$
becomes small so as to decrease $N_{\rm F}^0$. The decrease of $\mu$
also decreases $\Pi({\bf q}_{\rm max},0)$ because of $\partial
\Pi({\bf q}_{\rm max},0)/\partial\mu>0$; thus the factor
$1-U\Pi({\bf q}_{\rm max}\to{\bf q}_{\rm FF},0)$ is still finite (or
eq. (\ref{eq.15}) is not satisfied) at $h_{\rm FF}^{\rm MF}$ and the
phase transition to the FF state is avoided. Thus, the shift of the
chemical potential due to the superconducting fluctuations is crucial near
$h_{\rm FF}^{\rm MF}$, in contrast to the case of the ordinary weak-coupling BCS state.
\par
The above discussion is also applicable to the case of $\partial \Pi({\bf q}_{\rm max},0)/\partial\mu<0$ (negative $N_{\rm FL}$). In this case, in order to increase $N_{\rm F}^0$, $\mu$ becomes large and thus $\Pi({\bf q}_{\rm max},0)$ decreases. As a result, we obtain $1-U\Pi({\bf q}_{\rm max}\to{\bf q}_{\rm FF},0)>0$ at $h_{\rm FF}^{\rm MF}$ and the FF phase transition is again avoided.  
\par
We note that this mechanism also works in the region $h<h_{\rm FF}^{\rm MF}$ until $\Pi({\bf q},0)$ takes a maximum value at ${\bf q}=0$. Thus, the second-order phase transition into the FF state does not exist in the $h-T$ phase diagram. 
\par
\vskip2mm
{\it Summary and discussion--}
In conclusion, we have discussed the superconducting fluctuations in a three-dimensional superconductor under an applied magnetic field using the NSR theory. Under a strong magnetic field, the second-order phase
transition associated with the FF state has been examined within the
mean-field theory. However, this phase transition actually does not
occur at finite temperatures due to superconducting fluctuations
unless the effect of the crystal lattice is taken into account. We also
showed that, even if the pairing interaction is weak, the shift of the
chemical potential induced by the superconducting fluctuations is crucial for
understanding what actually occurs near $h_{\rm FF}^{\rm MF}$. This is
in contrast to the ordinary BCS case, where we can set
$\mu=\varepsilon_{\rm F}$ as far as the weak-coupling regime is considered.
\par
When the second-order phase transition associated with the FF state is absent, it is interesting to consider what sort of superconducting phase transition would be realized under the strong magnetic field. In this regard, it is known that within the mean field theory, when the FF state is neglected, the first-order superconducting phase transition occurs at the Clogston limit. In addition, the present discussion is within the second-order phase transition to the FF state and thus the possibility of the first-order phase transition is not denied. Since the present approach is valid only for $T\ge T_{\rm c}$, we must extend the present method to the superconducting state in order to discuss these first-order phase transitions. This problem remains for future discussion.
\par
Although real superconductors are more or less anisotropic due to the presence of a crystal lattice, we can expect spatially isotropic superfluidity in trapped gases of Fermi atoms such as $^{40}$K and $^6$Li.\cite{Jin,Granade,Ohashi2} In these systems, two hyperfine states corresponding to electrons with up- and down-spins feel different potentials in a magnetic trap potential. This situation is similar to superconductivity under magnetic field. In addition, the spatial isotropy can be controlled experimentally by introducing an optical lattice to trapped gases.\cite{Delmer} Thus, if superfluidity is realized in this system, it is expected that we can examine how the FF state would be stabilized when the anisotropy of the system is introduced.
\par
{\it Acknowledgements--}
The author would like to thank Dr. K. Samokhin for valuable discussions and useful comments on this topic. He also thanks Professors A. Griffin, S. Takada, H. Shimahara and Dr. T. Momoi for useful discussions. He was financially supported by a Japanese Overseas Research Fellowship.
%

\begin{figure}
\includegraphics[width=10cm,height=10cm]{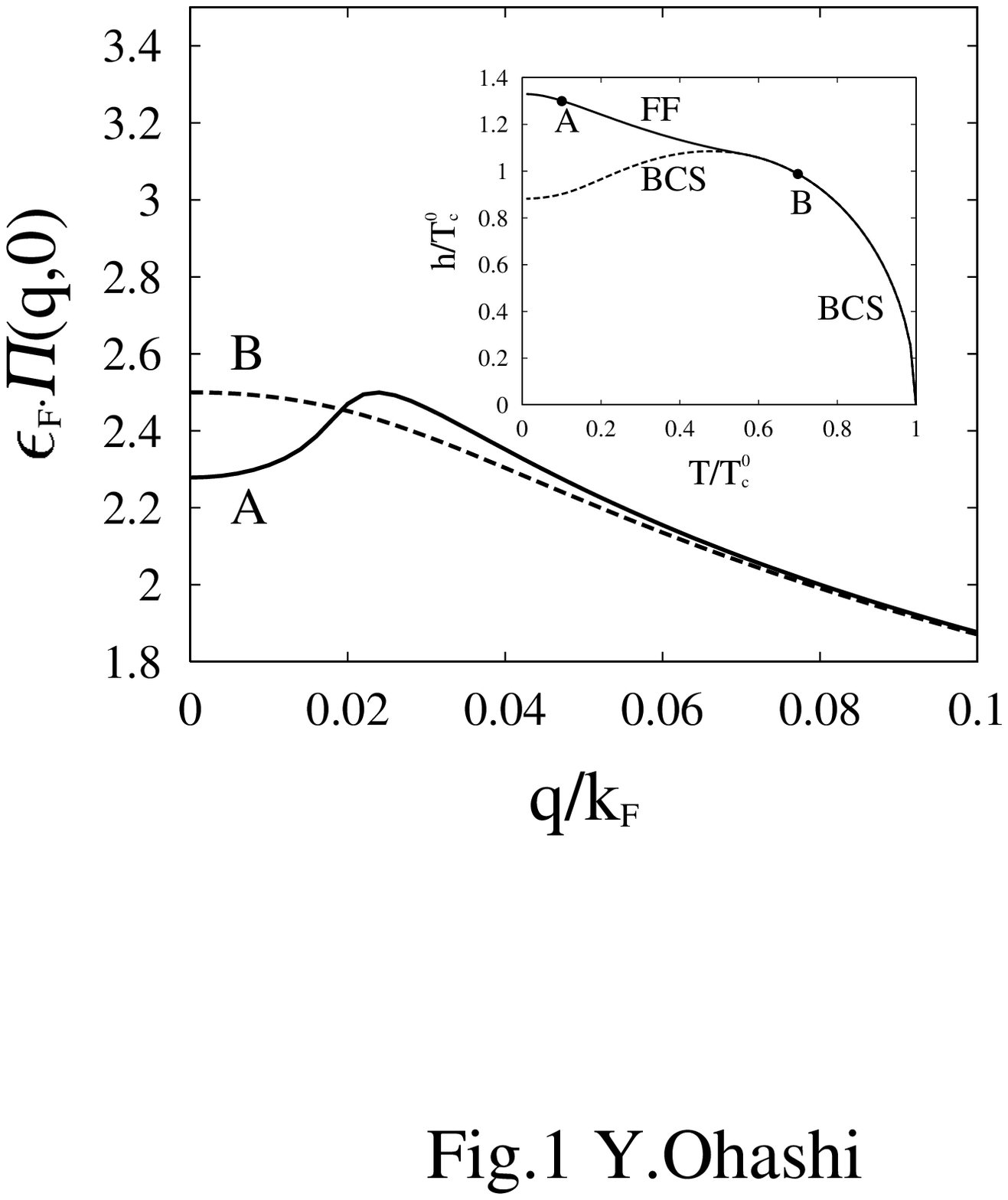}%
\caption{
$\Pi({\bf q},0$) as a function of $|{\bf q}|$ at the superconducting phase transition obtained within the mean-field theory (i.e., neglecting superconducting fluctuations). The solid line is at $h_{\rm FF}^{\rm MF}(T/T^0_{\rm c}=0.1$) ('A' in the inset) while the dashed line is at $h_{\rm FF}^{\rm MF}(T/T^0_{\rm c}=0.7)$ ('B' in the inset). Inset: $h-T$ phase diagram within the mean-field theory. The upper line is the FF phase transition while the lower one is the BCS phase transition under the assumption of second-order phase transition. We set $\mu=\varepsilon_{\rm F}$ and $U/\varepsilon_{\rm F}=0.4$. (Thus eq. (\ref{eq.15}) is satisfied at $\varepsilon_{\rm F}\Pi({\bf q},0)=2.5$.) We have introduced the Gaussian cutoff $e^{-(\epsilon_{\bf p}/D)^2}$ with $D=2\varepsilon_{\rm F}$ in solving eq. (\ref{eq.8}). The first-order transition (Clogston limit) and the phase boundary between the FF state and the BCS state are not shown.
\label{fig1} 
}
\end{figure}


\begin{thebibliography}{99}
\bibitem{note3} When the attractive interaction is so strong that preformed Cooper pairs can appear, ${\rm Im}\Pi({\bf q},z)$ has a finite gap ($\alpha=0$). We do not consider this strong coupling regime here.
\bibitem{note2} The author would like to thank Dr. K. Samokhin for pointing out the similarity between the FF state and smectic liquid crystal.
\end{thebibliography}
\end{document}